\documentclass[11pt,oneside]{article}

\usepackage[utf8]{inputenc} 
\usepackage[T1]{fontenc}    
\usepackage{amsmath}
\usepackage{amsfonts}
\usepackage{amsthm}
\usepackage{amssymb}
\usepackage{color}
\usepackage{hyperref}       
\usepackage{url}            
\usepackage{graphicx}
\usepackage[round]{natbib}
\usepackage{doi}

\usepackage[a4paper]{geometry}

\usepackage{mathabx}

\theoremstyle{definition}
\newtheorem{definition}{Definition}[section]
\newtheorem{example}[definition]{Example}

\theoremstyle{plain}
\newtheorem{theorem}[definition]{Theorem}
\newtheorem{proposition}[definition]{Proposition}

\newtheorem{lemma}[definition]{Lemma}

\theoremstyle{remark}
\newtheorem{remark}[definition]{Remark}

\def\endproof{}
\def\proof{\paragraph{Proof.}}
\def\proofof[#1]{\paragraph{Proof#1.}}

\newcommand{\R}{\mathbb{R}}
\newcommand{\F}{\mathcal{F}}
\newcommand{\N}{\mathbb{N}}
\newcommand{\Z}{\mathbb{Z}}
\renewcommand{\P}{\mathbb{P}}
\newcommand{\E}{\mathbb{E}}

\def\lnorm{\left\ldbrack}
\def\rnorm{\right\rdbrack}

\def\esssup{\mathop{\rm ess\,sup}}

\def\Span{\text{\rm span}\,}
\def\rank{\text{rank}\,}
\def\rk{\text{rank}\,}

\def\im{\text{im}\,}
\def\RR{\R} 
\def\NN{\N} 

\def\PCA{\text{PCA}}

\def\trace{\text{tr}}

\def\cle{\preceq}
\def\cll{\llcurly}
\def\clle{\stackrel\llcurly{\_\!\_}}
\def\cg{\succ}
\def\cge{\succeq}

\def\1{{\bf 1}}

\long\def\notiz#1{{\bf #1}}
\def\notiz#1{}

\def\GF{{\color{red}R}}
\def\GX{{\color{blue}G}}
\def\text#1{\mbox{\rm #1}}
\def\plusG{{\color{red} \dotplus}}
\def\dotsumG{\color{red}\sum\limits^{.}}
\def\plus{{\color{blue}\ddot+}}
\def\dotsum{\mathop{\color{blue}\sum\limits^{..}}}

\def\felix#1{}
\def\GF{R}
\def\GX{G}
\def\plusG{\dotplus}
\def\dotsumG{\sum\limits^{.}}
\def\plus{\ddot+}
\def\dotsum{\sum\limits^{..}}

\allowdisplaybreaks
\title{A semi-group  approach to Principal Component Analysis}
\author{Martin   Schlather\footnote{Institute of Mathematics,
    University of Mannheim, 68159 Mannheim, Germany. Email:
    schlather@math.uni-mannheim.de} {} and Felix
    Reinbott\footnote{Institut für Mathematische Stochastik,
      Otto-von-Guericke-Universit\"at Magdeburg, 39106 Magedeburg,
      Germany.}} 
\date{\today}

\begin{document}
\maketitle

  \begin{abstract}
    Principal Component Analysis (PCA)  is a well known procedure to reduce intrinsic complexity of
    a dataset, essentially through simplifying the covariance
    structure or the correlation structure.
    We introduce a novel algebraic, model-based
    point of view and provide in particular
    an extension of the PCA to
    distributions without second moments
    by formulating the PCA as a best
    low rank approximation problem.
    In contrast to  hitherto existing approaches,
    the approximation is based on a kind
    of spectral representation, and not on the real space.
    Nonetheless, the prominent role of the eigenvectors
    is here reduced to define the approximating surface and its
    maximal dimension. In this perspective, our approach is close to the original 
    idea of \cite{pearson01} and hence to autoencoders.
    Since variable selection in linear regression can
    be seen as a
    special case of our extension, our approach gives some
    insight,
    why the various variable selection methods, such as forward
    selection and best subset selection,
    cannot be expected to
    coincide.
      The linear regression model itself and the PCA regression appear as
    limit cases. 
\end{abstract}

{\small
	}{
	\phantom{\textit{2010 MSC}:}{Secondary 60B15, 60E07, 62H05, 62J05} 
}




\section{Introduction}

Principal Component Analysis (PCA),
introduced by \cite{pearson01}, has been one of the most commonly used statistical methods for reducing the complexity in datasets:
for $n$ samples with a high number of features $p$, a linear subspace is chosen
in favor of a simpler representation of the data.
Under the assumption of existing variances,
these applications have been justified by
many theoretical results and have been applied successfully in a wide
variety of scientific fields.
Overviews can be found in \citep{ HastieTrevor2009TEoS,
JolliffeIanT2002PCA, RingnerPCA}, for instance.
Some special care is needed when the distribution of the data deviates considerably
from the multivariate normal distribution. In particular, probability laws
without second moments or counting distributions may lack theoretical justification or
a good interpretation when PCA is applied \citep{JolliffeIanT2002PCA}.

Several generalizations have been considered \citep{Bengio2013,
  CandesRobustPCA, Hofmann_2008, Silverman96, vidal2016generalized}.
A particularly flexible one is the  autoencoder, a neural network architecture that
has been considered both in mathematics and statistical learning \citep{BALDI198953, JolliffeIanT2002PCA, OJA198569}.
The autoencoder is based on the formulation of a regression problem, where the data is reconstructed
by a map that simplifies the stucture in some sense.
Often, simplification is meant in terms of mapping into a lower dimensional space  and then back to the original space
of the data. Here, the admissible maps are possibly nonlinear and chosen such that a certain loss function is minimized.
Classic PCA appears as a special case of choosing
linear maps.

Algebra plays a dominant role in founding certain areas of probability theory,
in particular stochastic processes \citep{sasvari05,schilling2012bernstein, StrokorbSchlather2015}. 
When special problems are adressed, modern algebra has also found applications in statistics,
for instance, in design of experiments \citep{bailey2004association} or in learning
theory \citep{watanabe2009algebraic}. Surprisingly, when data themselves are considered,
 ‘linearity’ usually refers to vector spaces over the field of real numbers,
although
many random variables exhibit natural linearity with respect to other
operations \citep{Golan99, 1974Tssc}. An exception in extreme value
theory is \cite{Klueppelberg2021} who refer to the tropical algebra,
however in a different context than PCA.
Distributions with a stability to some given algebraic operation
typically stem from limit laws and hence
are infinitely divisible,  again not necessarily with respect to the
field of real numbers \citep{DavydovZuyevMolchanov2008}.
In our algebraic approach, the classic PCA reappears as the Gaussian
case.


A particular class of distributions without
guaranteed second moments, which exhibits linearity with respect to maxima instead of addition,
are the max-stable distributions~\citep{2006EVT, ResnickSidneyI1987EVRV,
2005Esia}.
In contrast to the multivariate Gaussian distribution,
the dependence structure of max-stable distributions
cannot be fully described by
bivariate
characteristics like the
covariance \citep{2004Soe, StrokorbSchlather2015}.
Hence, decomposing any such derived matrix cannot be sufficient, at least from
a theoretical point of view
\citep{JCW20}.
\felix{
Section \ref{sec:maxstable} deals exemplarily with the max-stable case.
ans shows that our approach
remains numerically treatable,
using classical
estimators from extreme value statistics
as well as algorithms from nonlinear optimization.}

We consider here also
intrinsically vector-valued data, which  appear in colour coding, for instance.
Special cases there of are matrix-valued data, which appear in a
single measurement, for example in functional magnetic resonance imaging \citep{WST15}.
Linear regression models can also be seen as a special case of
vector-valued data. Both variable selection and 
classic PCA are dimension reducing methods. The ideas are frequently
combined, leading to the PCA regression analysis
or the sparse PCA, for instance \citep{HastieTrevor2009TEoS}.
Nonetheless,  variable selection and PCA have been considered as
different methods.

Central part of the paper is the definition of the
generalized PCA in Section \ref{sec:sec.PCA}. It is based
on generalizations of several well-known notions, such as stable
distributions and quadratic variation (Section \ref{sec:foundations}).
An important specification of our approach is the PCA for extreme values
(Section \ref{sec:maxstable}).
Some background information
\felix{and some proofs are}%
is
given in the appendix.

\section{Foundations}\label{sec:foundations}
Since classic
PCA minimizes  the
mean square of the residuals \citep{pearson01}, calculating the
difference between random variables is implicitly required.
In our generalization towards extreme values
with Fr\'echet margins,
we replace the abelian group $(\RR, +)$ by the semi-group
$([0,\infty), \vee)$ where $a \vee b := \max\{a, b\}$.  Since the
calculation of a
difference is impossible in a semi-group context,
we provide a workaround for the mean square of the residuals, here.
First, we have to declare for which random vectors we have
a workaround (Section \ref{sec:multivariate}). Essentially,
these vectors have a stable
distribution
(Section \ref{sec:stable}).
In Subsection
\ref{sec:distance},
we define a convenient
distance between random vectors, which avoids the calculation of residuals.
This
semi-metric is based on a semi-scalar product (Subsection
\ref{sec:uncorrelated}), which itself is based
on a kind of valuation principle (Subsection \ref{sec:valuation}).
The latter is fundamental, since it (i)~generalizes the
quadratic variation, (ii) is unique in important cases and
(iii)~throws new light on the variance.



\subsection{Semigroups and Semirings}
Since semigroups are not that frequently used in a statistical
context, we repeat basic notions. See \cite{Golan99} for a general
introduction, for instance.
Throughout the paper, we will use $+$, $\vee$ ,$\plusG$, and $\plus$  for
the binary operator of the standard
addition, the maximum, a general semigroups and a general semiring, respectively.
The corresponding multi-operators are denoted by $\sum$, $\bigvee$,
$\dotsumG$ and $\dotsum$.

\begin{definition}
  Let $\GX$ be a nonempty set and $\plusG \colon \GX  \times \GX \to \GX$ be an associative operation, then the tuple $(\GX, \plusG )$ is called a semigroup.
A semigroup is called
  \begin{enumerate}
    \item a \emph{monoid} with identity element $0$, if an element
      $0\in \GX$ exists, such that $\alpha \plusG 0 = 0 \plusG \alpha = \alpha$ for all $\alpha \in \GX$.
    \item \emph{commutative}, if $\alpha_1 \plusG \alpha_2 = \alpha_2 \plusG \alpha_1$ for all $\alpha_1, \alpha_2 \in \GX$.
    \item \emph{topological}, if the set $\GX$ has a topology $\tau$, $(\GX, \tau)$ is a topological space and the map $\plusG \colon \GX \times \GX \to \GX, (\alpha_1, \alpha_2) \mapsto \alpha_1 \plusG \alpha_2$ is continuous.
  \end{enumerate}
\end{definition}

\begin{definition}
  A  set $(\GF,\plus, \cdot)$ with addition $\plus$ and
  multiplication $\cdot$ is called a
  semiring if
  \begin{enumerate}
    \item $(\GF, \plus )$ is a commutative monoid with identity element $0$,
    \item $(\GF, \cdot)$ is a monoid with identity element $1$,
    \item multiplication is left and right distributive, i.e.,
      $\alpha_1  (\alpha_2 \plus \alpha_3) = \alpha_1  \alpha_2
      \plus \alpha_1  \alpha_3$ and
      $ (\alpha_2 \plus \alpha_3)  \alpha_1 = \alpha_2 \alpha_1
      \plus \alpha_3 \alpha_1$
    \item $0 \cdot \alpha = \alpha \cdot 0  = 0$ for all $\alpha \in \GF$.
  \end{enumerate}
 \end{definition}

\begin{example}
  Examples of practically relevant semirings are, for instance,
  $(X, +, \cdot)$  with $X\in\{\R, \Z, \N_0,
[0, \infty)\}$,  $([0,\infty), \vee, \cdot)$, $(\RR, \vee, +)$, and
$(\RR^{d\times d}, +, *)$, where $*$ denotes the matrix multiplication.
If $M$ is a non-empty set, then $(2^M, \cup ,\cap)$ is also a semiring.
Last but not least, the quaternion number system is a semiring, which
is used in certain areas of
physics, see \cite{menannomazzotti12}, for instance.
\end{example}

Essentially, the definition of a semiring means that
the inverses with respect to addition and multiplication are missing
and that the multiplication is not necessarily commutative.
However, we will always assume that $\plusG$ is commutative,
$\GX$ and $\GF$ have at least two elements,
and that $\GX$ and $\GF$ are topological.
Since the focus of the paper is on algebraic aspects, we
assume for ease that both, $\GX$ and $\GF$, are Polish.
Further, we will drop the multiplication sign in
formulae whenever possible. On the other hand, especially when several
different semigroups or semirings are involved in a formula, we may clarify neutral
elements and operators with indices.

In our set-up, the ``scalar'' random variable takes values in a monoid
$\GX$. Much more structure will be imposed on the set $R$,
which indexes the distributions and which
will be a semiring. Primarily,
it is this index set $R$ that is extended to higher
dimensions, the so-called semimodule.

\begin{definition}
  Let $(S, \oplus)$ be a commutative semigroup and $(\GF, \plus,
  \cdot)$ a
  semiring.
  Let $\odot\colon \GF \times S \to S$ be a  mapping that satisfies for all
  $\alpha,\beta \in \GF$ and $x,y\in S$ the following properties:
  \begin{eqnarray*}
    & \alpha \odot (\beta \odot x) = (\alpha \beta) \odot x\\
    & \alpha \odot (x \oplus y) = (\alpha \odot x) \oplus (\alpha \odot y) \\
    & (\alpha \plus \beta) \odot x = (\alpha \odot x) \oplus (\beta \odot x)  \\
    & 1_\GF \odot x = x, 0_\GF \odot x = x \odot 0_S = 0_S .
  \end{eqnarray*}
  Then $S$ is called a semimodule  over $\GF$.
  A subset $V \subset S$ that obeys the above conditions is called a subsemimodule.
  If $(S,\oplus)$ is a topological semigroup and  $\odot$ is
  continuous, the semimodule  is called topological.
  We write $(S, \plus)$ and $\cdot$ instead of $\odot$, if
  $S$ is canonic, e.g., if $S=\GF^d$ for some $d\in \NN$.
\end{definition}

\begin{definition}\label{def:rank}
  Let $S$ be a semimodule and $B\subset S$.
  Let $\# B$ be the cardinality of $B$.  The value
  $$
  \rk S := \min\{ \# B :  S=\Span(B)\}
  $$
  is called the \emph{rank  of $S$}.
\end{definition}

Note that the span in the preceding definition is calculated according
to the semimodule operations.
Linear maps will play an essential role for the reconstruction of the points.


\begin{definition}
  Let $S_1$ and $ S_2$ be topological semimodules over the same semiring
  $(\GF,\plus, \cdot)$,
  with the commutative semigroups $(S_1, \oplus_{S_1})$ and $(S_2, \oplus_{S_2})$.
  A map $H \colon S_1 \to S_2$ satisfying the conditions
  \begin{eqnarray*}
    & H (\lambda x) = \lambda H(x) & \forall \lambda \in \GF, x \in S_1 \\
    & H(x \oplus_{S_1} y) = H(x) \oplus_{S_2} H(y) & \forall x,y \in S_1
  \end{eqnarray*}
  is called \emph{$\oplus_{S_1}-\oplus_{S_2}-$linear}.
  If  $S_1$ and $S_2$ are both canonical, then $H$ is called \emph{$\plus-$linear}.
\end{definition}

\begin{remark}
  If $S_1 = \GF^d$ and $S_2= \GF^p$, a  \emph{$\plus-$linear} map $H$ can
  always be represented by a matrix $H\in \GF^{p \times d}$ such that
  $$
  Hx := \Big(\dotsum_{j=1,\ldots, d} H_{ij} x_j\Big)_{i=1,\ldots, p},\qquad x \in \RR^d
  .
  $$
\end{remark}

Although the definitions above are in analogy to the definitions of a vector space and
a linear mapping, the consequences of the transition from groups to
semi-groups  are severe.
For instance, the dimension of a subspace of a finite dimensional
space does not necessarily exist.
Appendix \ref{sec:ranks}
gives some implications that are particularly important when dealing
with extreme values.
Appendix \ref{sec:ranks} also delivers implicitly
 arguments,
why a constructive approach via explicit multivariate
distributions is chosen
to define the generalized PCA, and not via an abstract formulation based
on subsemimodules or on the rank of a matrix.

\subsection{Stable distributions}\label{sec:stable}

For a general approach to PCA without existing variance we need a generalized notion of stable
distibutions, where we replace the standard addition by an arbitrary
semigroup operation. Some additional
care is needed with respect to the scaling properties of random variables.
The following definitions provide the structure to develop a useful
theory.

\begin{definition}\label{def:stable.dist}
  Let $(\GX, \plusG)$ be a topological
  monoid 
  and  $(\GF, \plus, \cdot)$
   a semiring with an additional binary operation $\circ:
  \GF \times \GF \rightarrow \GF$.
  Let $\F := \{ F_\mu : \mu \in \GF\}$ be a set of distributions over
  $\GX$ and
  $H_{\mu }:\GX \rightarrow \GX$, $\mu \in \GF$,  
  measurable
  maps,
  such that
  \begin{eqnarray}
     H_{1 } \nonumber
    &= &{\rm id}_{\GX}
    \\     \label{eq:semistable2}
    H_{0}& \equiv &0_{\GX} 
    \\\nonumber
  H_{\mu} (X_\nu) &\sim& F_{\mu \nu},\qquad   \mu,\nu\in \GF,  X_\nu
                         \sim F_\nu
    \\     
     \label{eq:semistable1}
    X_\mu  \plusG X_\nu& \sim& F_{\mu \circ \nu},
                               \qquad X_\mu \sim F_\mu, X_\nu\sim
                               F_\nu \text{ independent}.
    \end{eqnarray}
   Then, the set $\F$ is called a \emph{stable set of distributions}.
   We write briefly $\mu X_\nu$ instead of $H_\mu (X_\nu)$.
 \end{definition}

 The following definition ensures that  transformations
 of random vectors have still the required distribution,
 see Proposition \ref{prop:consistent} below.


\begin{definition}
  Let $\F$ be stable set of distributions
  where all $H_\mu$, $\mu \in \GF$, are linear, then $\F$ is called \emph{linear}.
\end{definition}

\begin{example} \label{ex:inf.div}
  In case of symmetric $\alpha-$stable distributions
  $S_\alpha(\sigma, 0, 0)$, $\alpha\in(0,2]$, we have for $\sigma,
  \tau, \mu\in \RR$ that
  \begin{eqnarray*}
    \GX\equiv \GF &= &\RR\\
     H_{\sigma}(x)        &= & \sigma x
    \\
    \sigma\circ \tau              & = &(\sigma^\alpha+ \tau^\alpha)^{1/\alpha}
                   .
  \end{eqnarray*}
  Hence,  the set of symmetric $\alpha-$stable distributions is stable and
  linear. Here,
  the  Gaussian case is  included as
  $S_2(\sigma, 0, 0) = \mathcal{N}(0, 2 \sigma^2)$, see  \cite{samorodnitskytaqqu}.
\end{example}

\begin{example}\label{ex:multigauss}
  Matrix-valued data
  can be considered as vector-valued with special constraints, which 
  are modelled as a subsemiring of the semiring of matrices.
  An example of such a subsemiring is a set of  block diagonal matrices with
  fixed block structure.
  Let us consider here
  vector-valued data with values in $\RR^k$,  $k\in \NN$.
  Let
  \begin{eqnarray*}
    (\GX,\plusG) &=& (\RR^{k}, +)\\
    (\GF,\plus, \cdot) &=& (\RR^{k\times k}, +, *),
  \end{eqnarray*}
  where  $*$ is the standard matrix
  multiplication.
  Then $G$ is an abelian group and $R$ is a non-commutative ring.
  We identify $\GF$ with the set of $\plus$-linear maps,
  i.e.
  $$
  H_A:\RR^k \rightarrow 
  \RR^k, x\mapsto Ax\qquad  \text{for }A \in \GF
  ,$$so that 
  $1_{\GF}$ is the identity matrix, for instance.
  The $F_A$, $A\in \GF$, are not distinct, since for any $A,B\in\GF$ we have
  \begin{eqnarray*}
    A X_1 \sim B X_1& \Leftrightarrow& A A^\top = B B^\top
  .
  \end{eqnarray*}
  Finally, denote by $M^{1/2}$ a (through some alphabetic ordering
  uniquely defined) square root of a positive semidefinite matrix $M$.
  Then, for $A,B,C\in \GF$ and two
  independent random matrices $ X_1, X_1'\sim F_1$  we have
  \begin{eqnarray*}
    A \circ B &=& (A A^\top \plus B B^\top)^{1/2}
 .   \end{eqnarray*}
 Thus, the set of $k$-variate Gaussian distributions is stable and
 linear.
\end{example}

\begin{example}\label{ex:linregr.1}
  Let $(R,\plus, \cdot) =(\RR^{k\times k}, +, *)$ for some $k\ge 1$ as
  in the previous example. We switch to the standard notation.
  Let
   $P\subset R$ be the subsemiring of $k\times k$  matrices $A$ of the form
  $$
  A =\left(
    \begin{array}{cc}
      A_\sigma & A_\beta \\
      0_{(k-1)\times 1} & A_\mu \1_{(k-1)\times (k-1)}
    \end{array}
  \right),
  $$
  where $A_\sigma \ge0$, $A_\mu\ge0$, $A_\beta\in \RR^{1\times
    (k-1)}$,
  $ \1_{(k-1)\times (k-1)}$ denotes the unity matrix,
  and
  $A_\mu=0$ shall imply $A_\sigma=0$ and $A_\beta=0$. 
  Let for $\ell=1,\ldots, k$
  \begin{eqnarray*}
    S_\ell &:= & \{ A \in P : A_{\beta,j}=0,  j\not=\ell
                  \},\qquad 
    \\
    L_\ell&:=& \Span(S_1,\ldots, S_\ell) = \{ A \in P : A_{\beta,j}=0 \text{ for  } j > \ell
  \}.
  \end{eqnarray*}
  Then, for $\ell=1,\ldots,k$, the sets $S_\ell$ and 
  $L_\ell$ are subsemirings of $P$.
  We interprete this set-up as a framework for linear regression models.
  Let  $X_{1}\ldots, X_{k-1}$ be the predictor variables
  which  are typically assumed to be independent of the error
  $\varepsilon$.  Since we aim to show later on, that variable selection in linear
  modelling is a special case of our PCA,
  we assume that $Z=(\varepsilon, X_1,\ldots, X_{k-1})$ has any
  multivariate Gaussian distribution.
  Let $A
  \in L_k$. Then, $A_{\sigma}$ equals the standard deviation of the
  error
  if $\varepsilon \sim {\cal N}(0,1)$.
 The first component of $AZ$ equals the dependent variable $y$, i.e.,
  \begin{eqnarray}
    \label{eq:y}
    y= (AZ)_{11} = \sum_{i=1}^{k-1} A_{\beta,i} X_{i} + A_\sigma
    \varepsilon
    .
  \end{eqnarray}
   The set $S_\ell$ corresponds to a simple linear
   regression model based on
   $X_{\ell-1}$, for $\ell\ge 2$.
   The set $L_\ell$, $\ell < k$, denotes the models, where
   only the predictor variables $X_i$, $1\le i< \ell$, are considered.
   In our example, the intercept, which is crucial in practice, is
   missing, for ease of theoretical reasoning.
   The 
   family of distributions $\F$ corresponding to the $AZ$, $A\in L_k$, is a stable set,
   if $Z$ consists of independent Gaussian components. I.e., it
   can be shown that one of the roots $(A A^\top \plus B
   B^\top)^{1/2}$ is in $L_k$ if $A$ and $B$ are. Unfortunately, this
   is a trivial case for variable selection. For a general distribution of $Z$,
   a representation of the linear model as a stable family is unkown.
   Fortunately, $R$ itself is rich enough so that the PCA can
   be  applied,  cf.~Example \ref{ex:linregr.2}.
  \end{example}

\begin{definition} \label{def:inf.div}
  Let $\F$ be a linear, stable set of distributions.
  If $1\in \GF$ allows the division by $n$  in the
  sense that
  $$
  \mathop{\bigcirc}\limits_{i=1,\ldots,n} \nu_n = \nu_n\circ \ldots
  \circ \nu_n = 1
  $$
  for some $\nu_n\in \GF$ and any
  $n\in\NN$, then $\F$ is called a set of \emph{infinitely
    divisible distributions}.
\end{definition}

\subsection{Multivariate  distributions}
\label{sec:multivariate}

Stable sets of multivariate distributions
are already covered by Definition \ref{def:stable.dist}.
Here, we consider an alternative, constructive definition for
a multivariate version of a stable distribution that is
tailored for our generalized PCA and avoids existence problems.
Recall that $\mu X_\nu$ is an abbreviation for $H_\mu(X_\nu)$.

\begin{definition}\label{def:multi}
  Let $\F$ a stable set of univariate distributions and $d, n \in \NN$.
  Let $\F_0^d$ be the set of distributions given by
  \begin{eqnarray}
    \label{eq:multi}
   \nu W 
 :=
  \dotsumG_{j=1,\ldots, n}
  \left(\nu_{1j}, \ldots,
    \nu_{dj}\right)^\top W_j
 :=
  \dotsumG_{j=1,\ldots, n}
  \left(\nu_{1j}W_j    , \ldots,
    \nu_{dj}W_j    \right)^\top 
  \end{eqnarray}
  for all
  $$
  \nu = 
  (\nu_{ij})_{
    i=1,\ldots, d; j=1,\ldots,n}\in \GF^{d\times n}
 , $$
 independent random variables $W_j \sim F_{\mu _j}$, $\mu_j\in \GF$, and
  $W := (W_1,\ldots, W_n)^\top$.
   We write
  $$
  \nu W\sim F^d_{\nu,\mu}
  $$
  with $\mu=(\mu_1,\ldots, \mu_n)$
.
  Let $\F^d$ be the weak closure of $\F_0^d$.
  Then, an element of $\F^d$ is called a \emph{$d$-variate
    $\F$-distribution}.

  We call $(\GX, \GF, \F^d)$ a \emph{multivariate model}.
 It is called \emph{linear} if $\F$ is 
  linear.
\end{definition}

Definition \ref{def:multi} ensures that the univariate
margins of $F\in \F_0^d$ are in $\F$.
A  linear  multivariate model ensures that 
the $p$-variate margins  of $F\in \F_0^d$ are in $\F_0^p$.

\begin{proposition}\label{prop:consistent}
  Let $(\GX, \GF, \F^d_0)$ be a linear multivariate model,
  $X=(X_1,\ldots X_d)\sim F^d_{\nu,\mu}$,  $\xi =(\xi_1,\ldots, \xi_p)
  \in \GF^{p\times d}$, and
  \begin{eqnarray*}\label{eq:xiX}
    \xi  X
    &:=& ( \xi_{k\cdot}  X)_{k=1,\ldots, p}
         \quad  \text{ with }\quad
         \xi_{k\cdot} X = \dotsumG_{i=1,\ldots, d} \xi_{ki} X_i      
         .
  \end{eqnarray*}
  Then
  \begin{eqnarray}\label{eq:xiX1}
    \xi X 
    \sim F_{\xi\nu,\mu}.
  \end{eqnarray}
\end{proposition}

\proof{
  \begin{eqnarray*}
    \xi_{k\cdot} X
    = \dotsumG_{i=1,\ldots,d} \xi_{ki}\left(
  \dotsumG _{j=1,\ldots, n}
    \nu_{ij} W_j\right)
    =
    \dotsumG_{i=1,\ldots,d}\;
  \dotsumG_{j=1,\ldots, n} \xi_{ki}
    \nu_{ij} W_j
    =
    \dotsumG_{j=1,\ldots, n}\Big(\mathop{\dotsum}\limits_{i=1,\ldots, n} \xi_{ki} \nu_{ij} \Big) W_j    
    .
 \end{eqnarray*}
 }

\begin{remark}
  Both Definition \ref{def:stable.dist} and Definition
  \ref{def:multi} can be generalized slightly, replacing $H_\mu$
  by $H_{\mu, \nu}$, which is then applied to random vectors $X_\nu\sim
  F_\nu$, only. Then,
  Equation \eqref{eq:multi} is rewritten as
  $$
  X = \dotsumG_{j=1,\ldots, n} \left(H_{\nu_{1j}, \mu _i}, \ldots,
    H_{\nu_{dj}, \mu _j}\right)^\top W_j
  := 
\Big( \dotsumG_{j=1,\ldots, n} H_{\nu_{i}, \mu _j} W_j \Big)_{i=1,\ldots, d}
  .$$
  Assume $\GX\subset \RR$ and all $F \in \F$ are
  continuous and distinct, i.e., $F_\mu \not = F_\nu$ for $\nu\not= \mu$.
  Then  $H_{\nu,\mu } =   F_\nu^{-1}(F_\mu (x))$ is a possible choice.
  Here, $ F_\nu^{-1}$ denotes the pseudoinverse of $F_\nu$.
  In practice, monotonously increasing maps $H_{\nu,\mu}$ are
  preferred so that the $H_{\nu,\mu }$ are essentially unique.
  Hence, the generalized map $H_{\nu,\mu }$ suggests that our approach
  so far is
  essentially restricted to continuous distributions.
  \\
  The set of  Gamma distributions $F_\mu$ with fixed scale parameter and
  arbitrary non-negative shape parameters $\mu\ge 0$ obeys this
  generalized framework, but fails to be a stable set.
\end{remark}

\subsection{Variation}\label{sec:valuation}
In classic PCA the mean square of the residuals is minimized.
From  a model-based perspective, this refers to minimizing the variance
of the residuals. In our general approach,
the existence of the variance is not guaranteed, so that
we have to
consider a general function that attaches value to
a residual. We wish to minimize the sum of these attached values,
but face the additional difficulty that the calculation of the
residuals would need additive inverses.

We call the function that attaches values to a random variable a variation,
in generalization to the quadratic variation of a Wiener process.
Due to property \eqref{eq:additive} below,
it might be interpreted  as the number of underlying independent variables.

\begin{definition}\label{def:indep.measure}
  Let $\F$ be a stable set of distributions.
  A continuous map
  $\lnorm \cdot \rnorm : \GF \rightarrow [0,\infty]$
  is called a \emph{variation},
  if the following conditions hold
  \begin{eqnarray} \label{eq:consistent}
    \lnorm \mu  \rnorm
    & = \lnorm \nu  \rnorm, \quad \text{if } F_\mu  =
    F_\nu \hspace*{1.035cm}
    &\text{(consistent)} \\ \label{eq:positive}
    \lnorm \mu  \rnorm
    & > 0,
      \quad \mu \in \GF\setminus\{0\}
      \hspace*{1.35cm}
    & \text{(positive)} \\
    \lnorm 0 \rnorm
    & = 0 \hspace*{3.2cm}
    & \text{(degenerate element)} \\
    \label{eq:additive}
    \lnorm \mu \circ \nu \rnorm
    & = \lnorm \mu \rnorm + \lnorm \nu \rnorm
    \quad \mu ,\nu \in \GF \quad
    &\text{(additive).}
  \end{eqnarray}
  We call a variation \emph{scale invariant} if, additionally,
  \begin{eqnarray}\label{eq:invariant}
    \lnorm \mu \nu  \rnorm = \lnorm \mu  \rnorm \lnorm \nu  \rnorm \quad \text{for all  } \mu ,\nu \in \GF
  .\end{eqnarray}

  We also write  $\lnorm X \rnorm$ and $\lnorm F \rnorm$ for $X\sim F$ and $F \in\F$.
\end{definition}

 \notiz{
  Condition~3 in Definition  \ref{def:indep.measure} can be seen as a
  generalization to the Pythagorean theorem, so that two
  random variables $X_\mu\sim F_\mu$ and $X_\nu \sim F_\nu$
  are called uncorrelated if $\|\|$.
}

\begin{remark}
  \begin{enumerate}
  \item If $\lnorm \cdot \rnorm$ is scale invariant, we have
      $\lnorm \mu X_\nu \rnorm = \lnorm \mu \rnorm \lnorm
      X_\nu \rnorm$, so that rescaling of all components
      with the same value will not change the
      outcome of a PCA, provided the sets $B$ and  $I$ in Definition \ref{def:PCA} are
      also scale invariant.
    \item The function $(\mu,\nu) \mapsto  \lnorm \mu \circ \nu\rnorm$
      is negative definite, as is any function
      of the form $(x,y) \mapsto f(x) + f(y)$.
      Furthermore, $e^{\lnorm\cdot\rnorm}$ is a semi-character
      on $(\GF, \circ)$ with the identity as involution \citep{BCR}.
    \end{enumerate}    
 \end{remark}

\begin{proposition}
  If $\lnorm \cdot \rnorm$ is  scale invariant, then the
  following properties hold:
  \begin{enumerate}
  \item $\lnorm 1_\GF\rnorm =1\in\RR$ for the
    neutral element $1_\GF$ of $(\GF,\cdot)$.
  \item $\GF$ is division free, i.e.,
    for all $\alpha_1, \alpha_2
    \in \GF$ with
    $\alpha_1 \cdot \alpha_2 = 0$ we have $\alpha_1 = 0$ or
    $\alpha_2 = 0$.
  \item Let $(\GF,\cdot) \subset (\RR,\cdot)$ be a non-trivial interval
    with standard topology.
    Then, $\lnorm \mu\rnorm = |\mu|^\alpha$ 
     for some unique $\alpha\in\RR\setminus\{0\}$.
     \end{enumerate} 
\end{proposition}
\proof{
  \begin{enumerate}
  \item Equality \eqref{eq:invariant} yields
    $\lnorm 1_\GF \rnorm = \lnorm 1_\GF  \rnorm ^2 $.
     The positivity of the variation excludes $\lnorm 1_\GF \rnorm
     =0$. Hence, $\lnorm 1_\GF \rnorm =1$.
 \item  $0 = \lnorm \mu \nu  \rnorm = \lnorm \mu  \rnorm \lnorm \nu
   \rnorm$ implies that  $\lnorm \mu  \rnorm =0$ or $\lnorm \nu
   \rnorm=0$. The positivity of the variation yields $\mu=0$ or $\nu=0$.
 \item
   Let $A = \{ \log(x) : x\in  \GF \cap (0,\infty)\}$.
   The function $\ell:A\rightarrow\RR$, $ x\mapsto \log \lnorm e^x \rnorm$ is
   well defined on some nontrivial interval that includes $0$
   and is continuous there.
   Since $\ell$ obeys Cauchy's functional equation we get
   $\lnorm \mu \rnorm = \mu^\alpha$ for $\mu \in \GF \cap (0,\infty)$
   and some $\alpha\in\RR$. Now, assume that
   $\GF\cap(-\infty,0)\not=\emptyset$.
   Then, Cauchy's functional equation delivers that
   $\lnorm \mu \rnorm = |\mu|^\beta$ for $\mu \in \GF \cap (-\infty,0)$
   and some $\beta\in\RR$.
   For $\mu \in \GF \cap [-1,0)$
   we have  $|\mu|, \mu^2 \in \GF$, so that
   $\lnorm \mu\rnorm^2 = \lnorm \mu^2 \rnorm  = \lnorm |\mu|^2 \rnorm
   = \lnorm |\mu| \rnorm^2$. 
      Hence $\alpha=\beta$.
   The additivity yields $|\mu \circ \nu|^\alpha = |\mu|^\alpha +
   |\nu|^\alpha$ with $\alpha\not =0$.
   Assume $\alpha =0$. Then the continuity of the variation yields
   $1 = |1 \circ 1|^0 \not = 1^0 + 1^0$ in contradiction to \eqref{eq:additive}.
   Now, assume that  $\lnorm \cdot \rnorm_\alpha: \mu \mapsto
   \mu^\alpha$ and  $\lnorm \cdot \rnorm_\beta:\mu 
   \mapsto \mu^\beta$
   are two scale invariant variations with
   $\alpha,\beta\in\RR\setminus\{0\}$. Then, for all $\mu \in \GF$,
   $$
   (1 + |\mu|^\alpha )^{1/\alpha} = |1_{\GF}  \circ \mu| = (1 + |\mu|^\beta )^{1/\beta}
  , $$
   so that $\alpha=\beta$.

  \end{enumerate}

 }

\begin{example} In the symmetric $\alpha$-stable case,
  the so-called 
  covariation norm $\| \cdot \|$ assigns
 the
  parameter $\sigma\in \GF:=[0,\infty)$ to $X \sim S_\alpha(\sigma, 0, 0)$ for
  $\alpha \in (1,2]$, i.e.,
  $ \| X \|_{\alpha} = \sigma
  $.
  It follows immediately from the properties of $S_\alpha(\sigma, 0,
  0)$, see \cite{samorodnitskytaqqu},  that $\| X
  \|_{\alpha}^{\alpha}$ satisfies the four properties 
 of a scale invariant
  variation, that is,
  $
    \lnorm \sigma \rnorm = \| X \|_{\alpha}^{\alpha}
 $. For centered Gaussian variables with $F_1 ={\cal N}(0,1)$, the variation
 equals indeed the variance.
\end{example}

 \begin{example}
    In case of the the stable set  of $k$-variate Gaussian distributions,
    see Example \ref{ex:multigauss},
   the variation might be defined as
   \begin{eqnarray*}
    \lnorm A \rnorm & := & \trace (A A^\top) = \sum_{i=1}^k \sum_{i=1}^k
                           A_{ij}^2, \qquad A \in \GF:= \RR^{k\times k}
                           .
   \end{eqnarray*}
   Then, $\GF$ is division free if and only if $k=1$.
  \end{example}

\subsection{Semi-scalar product}\label{sec:uncorrelated}
Property \eqref{eq:additive} of the variation gives reason 
to generalize the notion ``uncorrelated'' to random variables
without existing variance. The following definition is tailor-made for
scale-invariant variations.

\begin{definition}\label{def:scalarproduct}
  Let $(\GX, \GF,\F^d)$ be a multivariate model,
  $\lnorm \cdot \rnorm$  a variation and $\lnorm 1 \plus 1\rnorm \not =2$.
  Let  $X$ and $Y$ be random vectors such that their distribution
  and that of $X \plusG Y$ are in $\F^d$.
 For $\mu \in \GF^d$ let
  $\lnorm \mu  \rnorm$  be an extension of the variation to a vector $\mu$.
  Then,
  $$
  \langle X , Y \rangle := \frac{\lnorm X
  \plusG Y\rnorm - \lnorm X\rnorm - \lnorm Y\rnorm}{\lnorm 1 \plus 1\rnorm - 2}
  $$
  is called the \emph{semi-scalar product between $X$ and $Y$}.
  The vectors  $X$ and $Y$ are called \emph{uncorrelated (positively /
  negatively correlated}) if $
  \langle X , Y \rangle=0$ ($\ge0$ respectively $\le0$).
\end{definition}


\begin{remark}
  Given the definition of a variation in the univariate case,
  the definition of the variation of a vector is not clear cut.
  A convenient definition is
  \begin{eqnarray}
    \label{eq:multi.variation}
     \lnorm \mu \rnorm := \sum_{i=1}^d \lnorm \mu_i
  \rnorm,
  \end{eqnarray}
 as it ensures \eqref{eq:consistent}--\eqref{eq:additive}
  without further assumptions.
\end{remark}

\begin{example}
  Let $X$ be a standard Gaussian random variable
  and the variation of a vector be the sum of the variation of the components.
Then the random vectors  $(1, -1)^\top X$ and $(1,1)^\top X$ are
jointly multivariate Gaussian, fully dependent, but uncorrelated
according to Definition \ref{def:scalarproduct}.
Note that the standard notion of ``uncorrelated'' is defined only for
scalar random variables. The generalized definition still implies that
two jointly bivariate, scalar Gaussian random variables are uncorrelated if
and only if they are independent.
\end{example}

\begin{example}
  In the max-stable case the operator $\plus$ is the maximum, so that
  $1 \plus 1 =1$ and hence $\lnorm 1 \plus 1\rnorm  - 2 = -1 \not =
  0$.
  In the case of $\alpha$-stable distributions, however, the case $\alpha=1$ leads to
  $\lnorm 1 \plus 1\rnorm  - 2 = 0$, so that in particular the Cauchy
  distribution needs its own theoretical treatment or, at least, some
  limit considerations.
\end{example}

\begin{remark}
  Definition \ref{def:scalarproduct} suggests the interpretation that
  two random quantities are called uncorrelated if they behave as if they
  were independent. This behaviour has been made precise in terms of the
  variation.
\end{remark}

\begin{remark}
  Linearity of the multivariate model is not sufficient to have $X\plusG Y \sim F \in \F^d$
  in Definition \ref{def:scalarproduct}.
  As a well-known example, consider the set $\F$ of univariate Gaussian
  distributions with $H_\sigma (x) = \sigma x$, $\sigma\in\RR$. Let
  $X\sim {\cal N}(0,1)$ and $Y \sim \pi X$ where $\pi$ is a random
  sign, i.e., $\P(\pi = 1)= \P(\pi = -1)=0.5$. Then, $Y \sim {\cal N}(0,1)$, but
  the distribution of 
  $X+Y$ does not belong to $\F$.
\end{remark}

\begin{remark}
  For two jointly $\alpha$-stable, scalar random variables $X$ and
  $Y$ with scale parameter $\sigma$ and $\tau$, respectively, the codifference is
  defined as \citep{samorodnitskytaqqu}
  $$
  \tau_{X,Y} = \lnorm X \rnorm + \lnorm Y \rnorm - \lnorm
  X  -Y\rnorm
  $$
  and measures the difference between two variables. By way of contrast,
  $(\lnorm 1 \plus 1\rnorm - 2)\langle
  X, Y \rangle = \lnorm
  X \plusG Y\rnorm - (\lnorm X \rnorm + \lnorm Y \rnorm)$
  measures the difference in variation of a sum of two dependent variables
  and of two independent ones. Formally, $(\lnorm 1 \plus 1\rnorm - 2)\langle
  X, Y \rangle = -\tau_{X, - Y}$.
\end{remark}

\begin{lemma}
  Let $(\GX, \GF,\F^d)$ be a multivariate model,
  $\lnorm \cdot \rnorm$  a variation and $\lnorm 1 \plus 1\rnorm
  \not =2$.
  Let  $X$, $Y$ and $Z$ be random vectors such that their distributions
  and those of $X \plusG Y$, $X \plusG Z$ and $Z \plusG Y$  are in $\F^d$.
  Let $ \mu  \in \GF$.
  Then, the following assertions hold:
  \begin{eqnarray*}
    \langle X , X \rangle &= &\lnorm X\rnorm \ge 0
    \\
    \langle  X , Y  \rangle &= &\langle  Y , X  \rangle
   \\
    \langle X , 0 \rangle &=&0
    \\
    \langle  X \plusG Z , Y  \rangle  &=
    &\langle  X , Z \plusG Y  \rangle - \langle  X, Z  \rangle+
    \langle  Z, Y \rangle
    \\
    \langle X , X \rangle =0 &\Rightarrow & X \equiv 0,\text{ if the variation is scale invariant}
    \\
    \langle \mu  X , \mu Y  \rangle &= &\lnorm \mu \rnorm
    \langle  X , Y
    \rangle,\text{ if the variation is scale invariant}                             .
  \end{eqnarray*}
  If $X$ and $Y$ are independent, then they are uncorrelated.
 \end{lemma}

\subsection{Semi-metric between random vectors}
\label{sec:distance}

  The regression problem from classic PCA as given in
  \eqref{PCAregression} below is formulated using the
  squared $L^2$ distance, which is not a norm, but precisely fits the
  setting of a semi-metric $\rho$ 
   measures
  the gap between two random variables $X, Y$.

A semi-metric is given by the following three conditions
\begin{eqnarray}
  \rho(X, Y) & \geq 0 \quad & \text{(positivity)}, \label{d:0}\\
  \rho(X, X) & = 0 \quad & \text{(identity)}, \\
  \rho(X, Y) & = \rho(Y,X) \quad &
                                   \text{(symmetry)} \label{d.symmetry}
                                   .
\end{eqnarray}
With respect to the PCA we require further that $\rho$ is continuous and
\begin{eqnarray}
  \label{eq:rho.indep}
   \rho\Big(\sum_{i=1}^2 \nu_i X_i, \sum_{i=1}^2\xi_i X_i\Big) &=&
 \sum_{i=1}^2 \rho(\nu_i X_i , \xi_i X_i),\qquad\text{for }
                                                 \nu_i,\xi_i\in \GF
       \text{ and } X_1, X_2 \text{ indep.}\\
                 \rho(X, Y)& =& \rho(U, V) \qquad \text{for } X\equiv U
                                \text{ and }Y\equiv V \quad \text{(a.s.)}
                                \label{eq:rho.multi}
                                .
\end{eqnarray}

Proposition 3.2 in \cite{BCR} deals with the  generalization of a
squared difference of real values towards complex values, in the framework
of Hilbert spaces. The next definition carries over the implicit idea
given there.

\begin{definition}\label{def:semimetric}
  Let $(\GX, \GF, \F^d)$
  be a multivariate model, $\lnorm \cdot \rnorm$ giben by \eqref{eq:multi.variation}
  and $ \lnorm1\plus
  1\rnorm \not = 2$.
  For random vectors $X$ and $Y$ such that their distibutions
  and that of $X \plusG Y$ are in $\F^d$, let
  $$
  \rho (X,Y) := \lnorm X \rnorm + \lnorm Y\rnorm -  2 \langle X , Y \rangle 
  . $$
  Then $\rho$ is called the \emph{associated semi-metric}.
\end{definition}

\begin{lemma}
  Let $(\GX, \GF,\F^d)$ be a multivariate model,
  $\lnorm \cdot \rnorm$  a variation with $\lnorm 1 \plus 1\rnorm
  \not =2$,
  and $\rho$ be the associated semi-metric.
  Let  $X$ and $Y$  be random vectors such that their distributions
  and that of $X \plusG Y$  are in $\F^d$.
  Then, the following assertions hold:
  \begin{eqnarray}\label{eq:rho1}
     \rho(X,X)&=& 0\\\label{eq:rho2}
    \rho(X, 0) &=& \lnorm X\rnorm
   \\\label{eq:rho4}
      \rho (X,Y) &=& \lnorm X \plusG Y\rnorm -  \lnorm1\plus 1\rnorm
                     \langle X , Y \rangle
                     =
  \frac{
    \lnorm1\plus 1\rnorm ( \lnorm X\rnorm +  \lnorm Y\rnorm )
    -2  \lnorm X \plusG Y\rnorm }{\lnorm1 \plus 1\rnorm -2}
    \\\label{eq:rho3}
    \rho(X, Y) &=& \lnorm X\rnorm + \lnorm Y\rnorm =\lnorm X \plusG Y\rnorm , \text{ if $X$ and $Y$
                   are uncorrelated}
     \\\label{eq:rho5}\ \ \ \ 
     \rho(\mu X, \nu X) & > & 0, \text{ if $\mu X \not =\nu X$,
                             $(\GF,\plus) = (\RR, +)$,
                             $\lnorm \mu \rnorm = |\mu|^\alpha$, and
    $0<\alpha\not=1$}.
  \end{eqnarray}
  Furthermore, Equation \eqref{eq:rho.indep} holds.
  Now assume that the variation a vector is the sum of the variation
  of the components. Then,
  $\rho$ is well-defined on $\F_0^d$ if the representation of
  $X = \nu W \sim F_{\nu,\mu}\in \F_0^d$ is unique up to reordering of the summands.
  If $(\GF,\plus,\cdot)=(\RR,+,\cdot)$ and $\lnorm \xi \rnorm =
  \xi^2$ then $\rho$ is well-defined if the representation is unique
  up to orthonormal
  transformations $U$ with $UW\sim W$  i.e., $X =  (\nu U) (U^\top W)$.
\end{lemma}

\proof{
  Equalities \eqref{eq:rho1}-\eqref{eq:rho4} obviously hold.
  Inequality \eqref{eq:rho5} holds since $\mu X \not =\nu X$ implies
  $ X \not \equiv 0$ and then
  \begin{eqnarray*}
     \frac{
    \lnorm1\plus 1\rnorm ( \lnorm \mu \rnorm +  \lnorm \nu\rnorm )
    -2  \lnorm \mu \circ \nu \rnorm }{\lnorm1 \plus 1\rnorm -2}
    & = &
          | \mu|^\alpha
          \frac{
          2^\alpha ( 1+ | \xi|^\alpha )
          -2 ( 1 +|\xi|^\alpha) }{2^\alpha -2}    
  \end{eqnarray*}
  with $\xi = \nu / \mu$. The right hand side takes its unique minimum
  at $\xi=1$, which is $0$ due to \eqref{eq:rho1}.\
  Now, let $\GF = \RR$ and $\lnorm
  \xi \rnorm = \xi^2$. For any orthonormal matrix $U
  \in\RR^{n\times n}$ we have $X \equiv \nu  U U^\top W $.
  Denote by $\lnorm \nu \rnorm$ the sum of the variation of all
  components of a matrix $\nu$. Then,
  $$\lnorm \nu \rnorm = \trace
  (\nu \nu^\top) =  \trace (\nu U U^\top\nu^\top) = \lnorm \nu U
  \rnorm
  .$$
  Note that, by Maxwell's theorem \citep[Proposition 12.2]{kallenberg01},
 $U W\sim W$ holds for all orthonormal $U$
  if and only if the $W_j$ are centered Gaussian.
  
}

\section{Generalizing the classic PCA}\label{sec:sec.PCA}
In our model-based approach, the PCA is seen as an approximation
of a random vector $X$ with known distribution by some other random
vector $Y$ with a simpler
structure. The function that
maps a realization of $X$
to a realization of $Y$
is a projection in classic PCA.
This function is called a reconstruction function here.
We call a PCA inferable, if the existence and the knowledge
of the reconstruction
function is guaranteed.
We start with reviewing the classic PCA.

\subsection{Classic PCA and Autoencoders}
Classic PCA is usually understood as reducing the complexity  of
data 
in an optimal way with respect to the mean squared error. In general,
the data is assumed to be
an i.i.d. sample of $X=(X_1,\ldots, X_m)$.
Classic PCA
is  based on the solution of \citep{pearson01}
\begin{eqnarray}
  \label{PCAregression}
  \min_{H \colon \rank \thinspace H \leq p} \E \| X - H X \|_2^2.
\end{eqnarray}
It can readily be seen, that $H = V V^T$ is a solution to the
minimization problem, where  $V$ is the matrix of
the first $p$ eigenvectors \eqref{PCAregression}.
In particular, $H$ is a projection matrix, thus symmetric \citep{BALDI198953}.
In statistical literature, often $H$ is replaced by $V V^\top$  in
\eqref{PCAregression}, additionally assuming that $V$ is
orthonormal. To enforce uniqueness of the solution $V$ in the general
case, an ordering of the corresponding eigenvalues is further assumed.

This problem can be generalized as follows.
Let $L$ be a measurable loss function and  $\Theta$ an arbitrary parameter space with
elements $(\theta_1, \theta_2) \in \Theta$ used to parametrize two
measurable functions $f_{\theta_1}$ and $g_{\theta_2}$.
Then, the autoencoder problem is given as
\begin{eqnarray}
  \label{autoencoder}
  \min_{(\theta_1, \theta_2) \in \Theta}  \E  \bigl[ L(X , f_{\theta_1} \circ g_{\theta_2} (X)) \bigl] .
\end{eqnarray}
Under mild assumptions,  the existence of a
solution is guaranteed.

\begin{theorem}
  \label{existence_thm}
  Let $X$ be a random variable and $\Theta$ a compact metric space for the parameter $\theta$ of
  the reconstruction functions $r_{\theta} \colon \Omega' \to \Omega'$. Let
  $L$ be
  a loss function that is bounded from below by $0$ and continuous for any function of one fixed argument.
  For all $x \in \Omega'$, the map
  $r_{\cdot}(x) \colon \Theta \to \Omega'$
  be continuous. If for all $\theta \in \Theta$ it holds that
  \begin{eqnarray*}
    \| L(X, r_{\theta}(X)) \|_{L^{\infty}(\Omega, \mathcal{A}, \P)}
    := \esssup \bigl| L \bigl(X, r_{\theta} \bigl( X \bigl) \bigl) \bigl| < \infty,
  \end{eqnarray*}
  then a solution to the autoencoder regression problem
  \begin{eqnarray}
    \label{autoencoder_reconstr}
    \min_{\theta \in \Theta}  \E  \bigl[ L(X , r_{\theta} (X)) \bigl] .
  \end{eqnarray}
  exists.
\end{theorem}
\proof{
  The function $\E [ L(X, r_{\cdot}(X))] \colon \Theta \to [0, \infty)$ has by assumption compact preimage, thus
  it suffices to show that it is continuous. For arbitrary $\theta \in \Theta$ and any sequence $(\theta_n)_{n \in \N}$
  with limit $\theta$, we get by dominated convergence
  \begin{eqnarray*}
    \lim_{n \to \infty} \E \bigl[ L(X, r_{\theta_n}(X)) \bigl]
    = \E \bigl[ \lim_{n \to \infty} L(X, r_{\theta_n}(X)) \bigl]
    = \E \bigl[ L(X, r_{\theta}(X)) \bigl].
  \end{eqnarray*}
}

This means that under reasonable choices of the statistical model and
loss function
we always have a solution to the autoencoder problem.
We will go further in our approach and consider also
\begin{eqnarray*}
  \min_{\theta \in \Theta}  \E  \bigl[ L(X , Y_\theta) \bigl]
\end{eqnarray*}
for certain classes of random variables $\{ Y_\theta : \theta \in\Theta\}$.

\subsection{Generalized PCA}\label{sec:PCA}
  

Since the Hilbert space structure is given up here,
various generalizations of the classic PCA are thinkable.
We give four variants, which we consider particularly
interesting. Two notions directly correspond to variable
selection procedures in linear regression analysis.
The following definition is based on a general semi-metric,
although we have the associated semi-metric in mind,
since there is no proved evidence that our suggested
semi-metric should be preferred. The following definition allows that
the PCA does not have a solution.

\begin{definition}\label{def:PCA}
  Let $(\GX, \GF, \F^d)$
  be a linear multivariate model
  and $\rho$ be a semi-metric such that \eqref{d:0}-\eqref{eq:rho.multi} hold.
  Let $\F^d_0$ be given as in Definition \ref{def:multi}.
  Let   $p\in \NN$,
  $X=\nu W \sim F_{\nu,\mu}^d\in
  \F_0^d$ and $D((b_1,\ldots, b_p)) = \Span\{b_1,\ldots, b_p\}$
  for $b_i\in\GF^d$.
  For some closed $B \subset \GF^{d\times p}$
  and some subset $I(X) = I(\nu_1,\ldots, \nu_n) \subset \GF^{d\times n}$,
  the  \emph{$p$-variate B-I PCA}
  is
  defined as
  \begin{eqnarray}
    \label{eq:PCA}
    \PCA_p(X) =
    \mathop{\arg\min}_{b \in B}
    \inf_{\xi \in   D^n(b) \cap I(X)}
     \rho \left(X , \xi W \right)
  .\end{eqnarray}
  The PCA is called
  \begin{enumerate}
    \item \emph{exhaustive} if $B= \GF^{d\times p}$.
    \item \emph{forward} if
      \begin{eqnarray}
        \label{eq:sequ}
        B =\{ (b_1,\ldots,  b_p) \in \GF^{d \times p} : (b_1, \ldots, b_{p-1})
        \in \PCA_{p-1}(X)\}
        .
      \end{eqnarray}
    \item \emph{unrestricted} if $I=\GF^{d\times n}$.
    \item \emph{(linearly) inferable} if 
      \begin{eqnarray}\label{eq:I}
        I(X)  = I(\nu_1,\ldots, \nu_n) \subset
        \{ (H \nu_1,\ldots, H \nu _n)
        :  H \text{ a }(\plusG\text{-linear) map }
        \GF^d \rightarrow \GF^d
        \}.
      \end{eqnarray}
  \end{enumerate}

  A set of vectors $b_1,\ldots, b_p$ that is a solution
  to the $p$-variate PCA  is called a \emph{set of first $p$
  principal vectors for $F^d_{\nu,\mu} \in \F_0^d$}.
  Let $F^d_{n} \in \F_0^d$ and $b_{n,1},\ldots, b_{n,p}$ be
  corresponding sets of principal vectors. If
  $F^d_{n} \rightarrow_w F^d \in
  \F^d$ and $b_{n,i}\rightarrow b_i\in \GF^d$, then
  the
  set of vectors $b_1,\ldots, b_p$ is called  a \emph{set of first $p$
  principal vectors for $F^d \in \F^d$}.
\end{definition}

\begin{remark}
  Condition \eqref{eq:sequ} ensures that the principal vectors
  in forward PCA are in decreasing order of importance.
  If these principal vectors are orthogonal in a certain sense,
  they might be called eigenvectors. For instance,
  two vectors  $\mu,\nu \in \GF^d$ might be called orthogonal,
  if
  \begin{eqnarray}
    \label{eq:orthogonal}
    \langle\mu X, \nu X\rangle = 0 \quad\text{ for all } F \in \F
    \text { and }
    X \sim F
    .
  \end{eqnarray}
  Note that this is in general stronger than requiring $\langle\mu , \nu \rangle
  = 0 $.
  \\
  In the gaussian case, the vectors $\mu\in\RR^d $ and $\nu\in\RR^d$ are orthogonal
  in the sense of \eqref{eq:orthogonal}, if and only if
  they are orthogonal in the Euclidean sense.
  In the $k$-variate Gaussian case with $d=1$, two matrices
  $A,B\in\RR^{k\times k}$ are orthogonal if and only if $A B^\top=0$,
  i.e., if the rows of $A$ are all ortogonal to the rows of $B$.
\end{remark}

\begin{remark}  
  In some cases, it is sufficient to consider only $n=1$ in the definition
  of a multivariate distribution, e.g., when adding independent variables
  is not reasonable. Then, Definition \ref{def:PCA} still applies,
  if the operator $\circ$  and all conditions built on it are ignored.
\end{remark}

\begin{example}\label{ex:linregr.2}
  (Continuation of the linear regression model,  Example \ref{ex:linregr.1})
  Let $P$, $S_\ell$ and $L_\ell$ be defined as there, $d\ge 1$,  $1\le
  m \le \min\{k-1, d\}$, $1\le p \le k -m$,   and
  \begin{eqnarray*}
  S_\ell^d& = &\{ (A, \ldots, A)^\top \in P^d : A \in S_\ell\}
    \\
    S &=&  \bigcup_{\ell = m+1}^k S_\ell^d,
    \\
    B &=&  S^p, 
  \end{eqnarray*}
  Then the exhaustive $\PCA_p$ searches the best subset selection with up
  to $p$
  predictor variables for a linear regression model with a $d$-variate dependent
  variable, $k-m$ predictor variables, and $m$ error variables.
  The forward PCA performs the forward selection.
   \\
  Let us now consider some underlying structure of the 
  variable selection.
 Let $k\ge 3$ and $C, D\in L_k$ with $C_\sigma=C_\mu=C_{\beta,1} =D_\sigma=D_\mu=
 D_{\beta,2}=1$  and $C_{\beta,i}=D_{\beta,i}=0$, otherwise.
 Then, both equalities  $C + A D = D$ and $D + A C = C$
  are not solvable for $A\in L_k$. 
  We say that
  $R=L_k$ is not strictly preordered.
  Since Theorem~\ref{thm:2} of the Appendix is rather tight in its assumptions, which
  include strict preordering, we may expect that even
  the one-dimensional semimodule $L_k^1=L_k$ possesses subsemimodules with range larger than $1$.
  This is indeed the case, as  $C D \in S_\ell$
  for $C \in L_k$, $D\in S_\ell$ and
  $\ell\le k$. Then, $\Span\{S_{k-1}, S_{k}\}$ has rank 2, for instance.
  Assume, that  two matrices $C,D\in L_k$ are orthogonal in the sense
  of \eqref{eq:orthogonal}. Then, it follows that 
  either one of the corresponding linear regression models (i.e. the
  whole first
  line of the matrix) is identically $0$, 
  or  
  both linear models are
  deterministic, i.e., $C_{\sigma} =D_{\sigma}=0$,
  or both models are trivial, i.e., $C_{\beta} = D_{\beta} =0$.
  Hence, we will not be able to orthogonalize the vectors that span
  the subspaces of the exhaustive PCA and the forward PCA.
  Therefore, we may not
  expect that forward PCA and exhaustive PCA will be the same, 
  cf.\ Theorem \ref{ref:ortthogonal} below.
\end{example}

\begin{example}
    (Continuation of the linear regression model,  Example \ref{ex:linregr.1})
  Oher forms of  variable selection are possible.
  For instance, let the subsemiring be given by the matrices
  $$
  A =\left(
    \begin{array}{cc}
      A_\sigma & A_\beta \\
      0_{(k-1)\times 1} & A_X
    \end{array}
  \right),
  $$
  where $A_X$ is any matrix.
  For the $\PCA_p$ consider any matrix $A$ such that $A_\sigma=0$ and
  the last $k-1-p$
  lines of $A_X$ are all zero. Further, the matrix
  $(A_\beta^\top,       A_X^\top)
  $ shall have the same rank as $A_X$.
  This approach  balances out a good fit  of the dependent
  variable with a good fit of the predictor variables.
  Therefore, we might consider it as a ``PCA variable selection''.
  One extreme situation is that
  the variation puts nearly no weight on the 
  dependent variable. Then, we end up primarily with a PCA for the predictor
  variables, in other words, with the PCA regression \citep{JolliffeIanT2002PCA}.
  On the other hand, if  the variation puts  no weight on the predictor
  variables (condoning that Condition \eqref{eq:positive} is violated), already
  $\PCA_1$  delivers
  exactly the standard regression.
\end{example}

\begin{remark}\label{rem:barvinok}
  For the linearly inferable PCA, matrices $H$ might be considered,
  whose so-called Barvinok rank is at most $p$, i.e., the set~$I$
  in Defintion \ref{def:PCA}
  is given by means of all
  matrices $H$ of the form
  $H = H_1 H_2^\top$ with $H_1,H_2\in \GF^{d\times p}$.
  Then, we may reformulate the exhaustive, linearly inferable $\PCA_p$
  as
  $$
  \PCA_p(X) =
  \mathop{\arg\min}_{H_1,H_2\in \RR^{d\times p}}    \rho(X, H_1 H_2^\top X)
  . $$
  Hence, the optimization problem becomes a single $2 p d$-dimensional problem.
  A further advantage is that this
  choice follows closely the autoencoder idea.
  A disadvantage is, that the  Barvinok rank is rather restrictive, cf.\
  Appendix \ref{sec:matrixrang}.
\end{remark}

\begin{remark}\label{rem:fb}
  If the variation is scale invariant and $\rho$ is the associated
  semi-metric, then
  the exhaustive, unrestricted PCA reads
  $$
   \PCA_p(X) =
   \mathop{\arg\min}_{b \in B}
   \sum_{j=1}^n \lnorm \mu_j \rnorm
   $$
   with
   $$
   f_b(\nu_{\cdot j}),
   \qquad f_b(u) =   \min_{\xi \in   D(b)} 
  \frac{
    \lnorm1\plus 1\rnorm ( \lnorm u\rnorm +  \lnorm \xi\rnorm )
    -2  \lnorm u \plusG \xi\rnorm }{\lnorm1 \plus 1\rnorm -2},
  \quad
   u\in \GF^d
   .  $$
   That is,
   $$
   \PCA_p(X) =
   \mathop{\arg\min}_{b \in B}
   \int f_b(u) M(d u),
   \qquad
   M = \sum_{j=1}^n \lnorm \mu_j \rnorm  \delta_{\nu_{.j}}
   .  $$ 
\end{remark}


\begin{remark}
  Except for the linearly inferable PCA, the requirement of the
  linearity
  of the multivariate model
  seems to be excessive,  since  only the univariate margins of $X \plus
  \xi W$ enter in the associated semi-metric.
\end{remark}

\subsection{Coincidence of variants}
Since the four variants of a generalized PCA coincide in the Gaussian
case, we consider here general conditions for a coincidence in some
exemplary cases.

\begin{theorem}\label{ref:ortthogonal}
  Let the conditions of Definition \ref{def:PCA} hold
  with $\rho$ the associated semi-metric and scale invariant
  variation.
  Assume that for any subsemimodules $U\subset V \subset \GF^d$ and $m = \rank V
  - \rank U>0$, vectors $b_1,\ldots, b_m \in \GF^d$ exists such that
  $V = \Span( U, b_1,\ldots,  b_{m})$.  
  Assume that for any subsemimodule $V\subset \GF^d$ 
  and $\mu\in \GF^d$ a
  vector $\pi\in \GF^d$ exists with the following two properties:
  \begin{enumerate}
  \item $
  \Span(V,\pi) \supset \Span(V, \mu)
  $
  \item
  For all $\nu\in \Span(V,\mu)$ and $\xi\in \GF$
  a value $\theta = \theta(\nu, \pi, \xi)\in \GF$
  and a $\zeta\in V$ exists such that,
  for all $\eta\in \GF^d$ that are orthogonal to $\pi$
  in the sense of \eqref{eq:orthogonal},
  we have
  \begin{eqnarray}
    \label{eq:orth}
    \nu \plus \eta \plus \xi \pi  &=& \zeta \plus \eta \plus  \theta \pi
    \\
    \nu \plus \eta \text{ and } \theta \pi & \text{are} & \text{orthogonal}
  .\end{eqnarray}
  \end{enumerate}
 Then the unrestricted, forward PCA coincides with the
  unrestricted, exhaustive PCA.
\\
  If equality holds in \eqref{eq:I} and 
 $\pi$ has always a representation of the form
  $
  \pi = \nu^* \plusG \xi^* \mu
  $
  with $\nu^* \in V$ and $\xi^*\in \GF$, then
  the linearly inferable, forward PCA coincides with the
  linearly inferable, unrestricted PCA.
\end{theorem}
\proof{
  Condition \eqref{eq:orth} ensures that
  a sequence of principal vectors
  $b_1,\ldots,b_p$ can be
  replaced by a sequence of pairwise orthogonal vectors $b_1^0, b_2^0\ldots,
  b_p^0$, so that
  $\Span(b_1,\ldots, b_k) \subset \Span(b_1^0,b_2^0,\ldots,
  b_k^0)$ for all $k=2,\ldots, p$.
  Let $M_k^s$ be the minimum in \eqref{eq:PCA}
  for $D^n := U_k := \Span(b_1^0,\ldots, b_k^0)$.
  Then $M_k^s$ is decreasing and
  $
  M_k^s - M_{k-1}^s
  $
  only depends on $b_k^0$.
  Now, consider the unrestricted exhaustive PCA.
  Let $V_k$ be the space spanned by the principal vectors
  $c_1,\ldots, c_k$.
  Condition \eqref{eq:orth} ensures that pairwise orthogonal
  $c_1^0,\ldots, c_p^0$
  of $V_p$ exists such
  that either $c_i^0 = b_i^0$ or $c_i^0$ is orthogonal to
  $U_p$ in the sense of \eqref{eq:orthogonal}.
  We may assume further, that those vectors that are orthogonal to $U$ are ordered according to
  the rules of a forward PCA.
  Let $M_k^c$ be the minimum in \eqref{eq:PCA}
  for $D^n := \Span(c_1^0,\ldots, c_k^0)$.
  Again, $M_k^c$ is decreasing and
  $
  M_k^c - M_{k-1}^c
  $
  only depends on $c_k^0$. Let $M_0^c := M_0^s :=0$.
  Assume the two sequences $(M_k^c - M_{k-1}^c)_{k=1}^p$ and $(M_k^s -
  M_{k-1}^s)_{k=1}^p$ are identical. Let the spaces $V_i$ and $U_i$ be identical up
  to $k-1$, but $V_{k} \not = U_{k}$. Then
  $M_{k}^s - M_{k-1}^s = M_{i}^s-M_{i-1}^s$ for $i=k+1,\ldots, p$.
  This can be seen as follows. Let $i$ smallest such that
  $M_{k}^s - M_{k-1}^s > M_{i}^s-M_{i-1}^s$.
  As $V_{k} \not= U_k$ a vector $v\in V_i$ exists that is orthogonal to
  $U_p$.
  Replacing $b_i^0$ by $c_k^0$ gives a contradiction to $b_k$ being an
  optimal choice.
  Now assume that the sequences are different. Then $k\in\{1,\ldots,
  p\}$ exists such that  $M_{k}^s - M_{k-1}^s < M_{k}^c-M_{k-1}^c$.
  Replacing $b_k$ by $c_k$ leads to a contradiction as before.
  In case of linearly inferable PCA, the additional condition ensures
  that $\pi$ is can be represented by means if a $\plus$-linear map if $\nu$ and $\mu$
  do so. Except that the proof is the same as in the unrestricted case.
}

Consider the following counterexample for $(\GX, \vee, \cdot)$.
Let $X = (2,1)^\top W_1 \vee(1,2)^\top W_2$ with
iid.\
$W_i\sim F_1$.
A principal vector for the unrestricted PCA with $k=1$
cannot be a multiple of any of the two unit
vectors.
Hence, the
forward, unrestricted PCA for $k=2$ is
different from the exhaustive one.

The coincidence of forward PCA with exhaustive PCA in the above theorem needs
structural assumptions on the space $\GF^d$, essentially the existence
of a genuine scalar product.
The coincidence of the unrestricted PCA with the linearly inferable PCA needs
assumptions on the distribution. A trivial sufficient condition is
that a $\plus$-linear map $A$ (depending on $X$ itself) exist such that
$(W_1,\ldots, W_n)^\top = A X$.

\felix{\subsection{On a generalization of the linearly inferable
    PCA}to do?}

\def\Fr{F^{(\alpha)}}

\section{Example: max-stable  distributions}\label{sec:maxstable}
For $\alpha >0$, the univariate $\alpha$-Fr\'echet margins $\Fr_\lambda$, $\lambda >
0$, is given by
$$
\Fr_{\lambda}= \1_{(0,\infty)}(x) e^{-(\lambda / x)^\alpha}, \quad x\in\RR
.$$
\cite{2005Esia} call $\lambda$ the scale coefficient and define $\| X
\|_\alpha := \lambda$ for $X \sim \Fr_\lambda$.
For two independent random variables $X_\lambda\sim
\Fr_\lambda$ and $X_\mu\sim
\Fr_\mu$  we have $X_\lambda \vee X_\mu \sim \Fr_{\lambda \circ \mu}$
with $\lambda \circ \mu = (\lambda^\alpha + \mu^\alpha)^{1/\alpha}$.
Hence, $\lnorm \lambda \rnorm = \lambda^\alpha = \| X
\|_\alpha^\alpha$. Proposition 2.6 in \cite{2005Esia} shows that the associated
semi-metric $\rho$ is indeed a metric.
Let $\Fr_0(x) = \1_{[0,\infty)}(x)$. Then 
$\nu X_\lambda \vee \mu X_\lambda = (\nu \vee
\mu) X_\lambda$ and $\nu( \lambda \circ \mu) = \nu\lambda \circ
\nu\mu$ for $\lambda,\mu,\nu \ge0$.
Hence, $(\GX, \GF, \F^d)$ is a linear multivariate model
for $(\GX, \vee)$ and $(\GF, \vee, \cdot)$ with $G=R=[0,\infty)$.
In the light of \cite{2005Esia}, Definition \ref{def:multi} deals
with an extremal integral over a discrete measure with finite support, where
the typically infinite index set $T$ of a stochastic process is replaced by a finite one,
$T=\{1,\ldots,d\}$.

For ease, consider henceforth the case $\alpha =1$ only.
Then, the set of all $d$-variate max-stable distributions $F$ with Fr\'echet
margins are given by \citep{2005Esia, 2004Soe}
$$
-\log F(x) = \int_{\Xi} \bigvee_{i=1}^d\frac{u_i}{x_i}\, S(
du)
,\qquad x =(x_1,\ldots, x_d)\in(0,\infty)^d
,$$
where $\Xi  := \{ u \in [0, \infty)^d : \| u\|_q=1\}$ for some $q\ge
1$
and the so-called spectral measure $S$ satisfies
$$
\int_\Xi u_i S(d u) =\lambda_i
,\qquad i=1,\ldots, d.
$$
Here, $\lambda_i>0$ is the scale coefficient of the $i$-th componenent.
Let $X = \nu  W \sim \Fr_{\nu, \mu} \in \F_0^d$.
Without loss of generality let $\bigvee_{i=1}^d v_{ij} = 1$ for all
$j\in\{1,\ldots, n\}$,
i.e., $\|\nu_j\|_\infty = 1$ for $\nu_j = (\nu_{1j},\ldots, \nu_{dj})$.
Then, for $x\in(0,\infty)^d$, we have
\begin{eqnarray*}
  \P(X \le x)
    &=& \P( \nu_{ij}  W_j \le x_i, i=1,\ldots, d; j=1,\ldots, n)
  \\
    &=& \exp\left(-\sum_{j=1}^n \bigvee_{i=1}^d        \frac{\nu_{ij}}{x_i} \mu_j \right)
                =\exp\left(-\int_\Xi \bigvee_{i=1}^d        \frac{u_i}{x_i}
                S_\nu (d
                u) \right)
\end{eqnarray*}
with $$S_\nu = \sum_{j=1}^n \|\nu_j\|_q \mu_j \delta_{\nu_j /
  \|\nu_j\|_q}
,$$
such that
$$
\lambda_i = \int_\Xi u_i S_\nu(d u) =  \sum_{j=1}^n  \nu_{ij} \mu_j
,\qquad i=1,\ldots, d
.$$
We get for $f_b$ in Remark \ref{rem:fb}
$$
 f_b(u) = \min_{\xi \in   D(b)}( 2  \lnorm u \plusG \xi\rnorm 
- \lnorm u\rnorm -  \lnorm \xi\rnorm )
.$$
Let $\nu W \sim \Fr_{\nu,\mu}$.
If the variation of vector is defined as the sum of the variation of
the components, then  we have
$\lnorm \nu W \rnorm = \sum_{i=1}^d \sum_{j=1}^ n \nu_{ij} \mu_j =
\sum_{i=1}^d \lambda_i < \infty$
and
$$
f_b(u) =\min_{\xi \in D(b)} \| u - \xi\|_1
.$$
As  $f_{cb} = f_b$ for $c>0$ and $((0,\infty), \cdot)$ is a group,
 the exhaustive, unrestricted PCA reads 
$$
\PCA_p(X) =
\mathop{\arg\min}_{b \in B, \| b \|_q \le 1}
\int_\Xi f_b(u) S_\nu (d u)
.  $$
Since $f_b(u) \le \|u\|_1$ and hence
$0\le \int_\Xi f_b(u) S_\nu (d u) \le \sum_{i=1}^d \lambda_i$, the solution to the
PCA problem is not empty, cf.\ Theorem \ref{existence_thm}. Assume $S_{\nu_n} \rightarrow S_\infty$ in the
weak topology. Let $b_n\in B$ with $\|b_n\|_q \le1$ be  respective solutions.
Since  $\{ b \in B : \|b\|_q \le1\}$ is a 
compact set, we have 
$b_{n_i}\rightarrow b_\infty\in B$ with $\|b_\infty \|_q \le 1$ for a subsequence $(n_i)$.
Now, $\Xi$ is also a 
compact set
and $f_b(u)$ is continuous in $u$ and $b$.
Then, the function $(b, u) \mapsto f_b(u)$ is uniformly continuous
and a constant $C\ge 0$ exists such that
\begin{eqnarray*}
 \int_\Xi f_{b_{n_i}}(u) S_{\nu_i} (d u)
&\rightarrow& C
\qquad(i\rightarrow \infty)
.\end{eqnarray*}
We get $       \int_\Xi f_{b_{\infty}}(u) S_{\infty} (d u) =C$
by standard arguments.
Similarly, $$\mathop{\arg\min}_{b \in B, \| b \|_q \le 1}
\int_\Xi f_b(u) S_\infty  (d u)\ge C
.$$
Hence, for any spectral measure $S$, 
the exhaustive, unrestricted PCA problem equals
$$
\PCA_p(X) =
\mathop{\arg\min}_{b \in B, \| b \|_q \le 1}
\int_\Xi f_b(u) S (d u),
.  $$

The alternative choice
$\lnorm u  W_\xi\rnorm := \|u \|_\infty \xi$ for $u \in \GF^d$
and $W_\xi\sim F_\xi$,
looks appealing, since then  the variation yields the extremal coefficient,
$$
\lnorm \nu W \rnorm = \sum_{j=1}^n\mu_j  \bigvee_{i=1}^d \nu_{ij} 
= \int_\Xi \bigvee_{i=1}^d u_i\,  S_\nu(du)
,$$
but then
$$
f_b(u) =\min_{\xi \in D(b)} |\; \|u\|_\infty  - \|\xi\|_\infty\;|
$$
becomes pathological.

\section{Discussion}

120 years ago, \citep{pearson01} introduced the PCA to simplify the
structure of  data. At that time, statistical models were unknown.
Surprisingly, up to date, not much attempt has been made to
substantiate the PCA by some theoretical framework.
With our original aim to define  a PCA rigorously for extreme values
we suggest here a general model, where the classic PCA
reappears as the Gaussian case.
 Knowing from this paper, that
only a narrow family of distributions suits for construction,
it is even more challenging to understand theoretically, why PCA is that
successful, in particular, when the assumptions are not satisfied.

Linear regression models fit our framework and the PCA delivers the theoretical
basis for variable selection. Three points are remarkable. First, variable
selection splits up into two parts (as our PCA does in general): (i) estimation of
the underlying model; (ii) approximation of the estimated model by a simpler model.
Second, the predictor variables together with a scalar dependent
variable build a single vector-valued quantity. Hence, the PCA acts
on a formally univariate quantity, and the approximation
takes  place on a complex subspace structure of a
one-dimensional space. From this point of view, it is not surprising,
when the results of the various variable selection procedures are
difficult to compare.
Third, the difference between the error term and a scalar dependent variable
vanishes formally, as the latter takes the place of the first, see
Equality \eqref{eq:y}.

The set $B$ in the Definition \ref{def:PCA} of the generalized PCA
corresponds to the hidden layer of a standard
autoencoder. Two generalizations are introduced in Definition
\ref{def:PCA}.
The set $I$ refers to modifications of a standard autoencoder that are
common to neural networks.
The genuinely novel idea that is introduced into neural network
modeling is the following, though.
The sum that enters the (sigmoid) activation function is replaced 
by some other semigroup operation.
From a modelling point of view for neural networks,
this paper can be seen as a guideline for choosing such semigroup
operations, in particular in heterogeneous networks
\citep{piras1996heterogeneous}.
Indeed, recent papers show that the latter can be much more efficient
than homogeneous networks \citep{she2021heterogeneous},
in particular in a complex, space-time context
\citep{christou2019hybrid}.
For instance, the brain seems to consist of
various classes of neurons, which
process data markedly differently
\citep{stefanescu2008low,schliebs2009integrated}.

Classic PCA and many extensions thereof approximate data
by a lower dimensional hyperplane in the real space. In our approach,
we define a  multivariate distribution by means of  a  spectral
representation. Hence, our generalized PCA can be interpreted as
searching a best ``hyperplane'' approximation in a kind of Fourier
space. 
This contrasts in particular the approaches so far for extreme values
\citep{JCW20,DreesSabourin2021}.
The so-called spectral PCA \citep{thornhill2002spectral} is somehow close
to our approach,
but sticks to the idea of classic PCA: data of
stationary processes are first  Fourier transformed, before they enter the classic PCA.

The foundation of our generalized PCA on semi-groups made 
a workaround for the mean squares of the residuals  in classic PCA
necessary. As a side effect, hitherto existing approaches can be seen
in a different light.
For instance, the scalar product for random variables includes
operations from three different semi-groups, which coincide in the
standard (i.e. Gaussian) case. Further, scaling
parameters carry the structural information, not the random variables.
Finally, the variance is a theoretically convenient quantity, but
lacks an intuitive statistical interpretation --- being the square of
a nicely interpretable quantity is not satisfying. The variation
of Definition \ref{def:indep.measure},
however, has a nice practical interpretation, as it measures the amount
of randomness. The variance is now the (unique) specification of the variation
in case of Gaussian random variables.

Our approach includes various open questions. Obvious ones
are the uniqueness of the PCA outside trivially equivalent
representations,
the specification of the inferable PCA,
and the computation in reasonable time.
Finally, a slight extension of the current approach will be needed 
to perform a PCA that
is adapted to discrete distributions.

\section*{Acknowledgment}
   The authors are grateful to Khadija
      Larabi, Christopher D\"orr, Alexander Freudenberg, Sebastian
      Herzog,
      Ilya Molchanov,
      and
      Marco Oesting,
      for many helful
      comments.
      The
      authors gratefully acknowledge support by Deutsche
      Forschungsgemeinschaft through the Research Training Groups RTG
      1953 and RTG 2297.

\begin{appendix}
  \section{An algebraic excursion: ranks}\label{sec:ranks}
The change from spaces based on a field to spaces based on a semiring
is drastic. For instance, the dimension of a semimodule  is generally not
well defined, i.e., bases with different cardinalities might span the
same space
\citep{Golan99}. Therefore, the rank, i.e.
the minimal cardinality of all
sets of vectors that span the space,  is considered, see Definition \ref{def:rank}.
Furthermore, the rank of a subsemimodule might be greater than
the rank of the semimodule  itself.
This may happen even if the
rank of the semimodule  equals one.
A simple example is given by the semiring $(\NN_0, +, \cdot)$
and the subsemimodule $\NN_0\setminus\{1,2,4,7\}$ over $\NN_0$,   spanned by
the one-dimensional vectors $3$ and $5$.
The unique rank of a matrix in a Hilbert space context splits  up into
various definitions in a semiring context (Appendix \ref{sec:matrixrang}).


\begin{definition}
  A semiring $(\GF,\plus, \cdot)$ is called a \emph{semifield}
  if $\GF$ is division free and $(\GF,\cdot)$ is commutative.
  A semifield $(\GF,\plus, \cdot)$ is called
  \emph{proper}, if  $(\GF,\plus)$ is a group.
\end{definition}

Various definitions of the notion ``independence'', hence of a basis of a
semimodule, are given in literature \citep[Def 2.2.5]{gondran1984linear}.
We follow here Definition 2.2.2 in \cite{wagneur91}.




\begin{definition}
  \label{def:indep}
  Let $S$ be a semimodule  over a semiring 
  $(\GF, \plus , \cdot)$. A
  set of vectors $\{x_i\in S : i \in I \}$ indexed by the set $I$,
  is called \emph{independent} (or \emph{non-quasi-redundant}),
  if for all finite subsets
  $I_0\subset I$, all $k \in I_0$ and all $\alpha_i \in \GF$ and
  $\gamma \in \GF\setminus\{0\}$ we have
  \begin{eqnarray}\label{eq:indep.1}
    \dotsum_{i \in I_0\setminus\{k\}} \alpha_i x_i \not= \gamma  x_k
  .  \end{eqnarray}
\end{definition}

In case $(\GF, \cdot)$ is an abelian group,
$\gamma$ in condition \eqref{eq:indep.1} can be chosen to be $1$,
i.e., condition  \eqref{eq:indep.1}  is equivalent to
\begin{eqnarray}\label{eq:indep.1B}
  \dotsum_{i \in I_0\setminus\{k\}} \alpha_i x_i \not=  x_k
  ,  \end{eqnarray}
cf.\ \cite{gondran1984linear}.

\begin{definition}
  Let $S$ be a semimodule  over a semiring $\GF$.
  If $B\subset S$ is a set of independent vectors and if
  all $x\in S$ have a  representation
  $$
  x = \dotsum_{b\in B} \alpha_b b,\qquad \alpha_b\in \GF
  ,$$
  then $B$ is called a \emph{(semi-)basis of $S$}.
  If, for all bases $B$ of $S$, we have $\# B =\rk S$, then
  $\rk S$ is called the \emph{dimension of $S$}.
\end{definition}

\subsection{Rank of a subsemimodule}\label{sec:curse}

Exemplarily and in particular with respect to the PCA for extreme values,
we show here
that we
have a radical change of the behaviour of random vectors
from dimension $d=2$ to dimension $d=3$ if we consider a genuine
semiring instead of a  field. 

\begin{definition} 
  \label{def:inferior} Let $\GX$ be a semigroup and $\alpha,\beta\in\GF$.
  We write $\alpha   \cle \beta $,
  if $\gamma \in \GX$ exists, such that $\alpha  \plus
  \gamma  = \beta $.
   We call $\GX$  \emph{(canonically) preordered} if for all $\alpha ,\beta \in \GX$
  we have $\alpha  \cle \beta $ or $\beta \cle \alpha $.
\end{definition}


\begin{definition} 
  \label{def:strictinferior}
  Let $\GF$ be a semiring and $\alpha,\beta\in\GF$.  We write $\alpha
  \clle \beta $,
  if $\gamma \in \GF$ exists, such that $\alpha  \plus \gamma \beta  = \beta $.
  We call $\GF$ \emph{strictly preordered} if for all $\alpha ,\beta \in \GF$
  we have $\alpha  \clle \beta $ or $\beta \clle \alpha $.
\end{definition}

  \label{sec:proof.curse}

  \begin{lemma}\label{lemma:2}
    Let $\GF$ be a strictly preordered semiring, $S = \GF^2$
    and $x_{i}=(x_{i1}, x_{i2}) \in S$ for
    $i\in I:=\{1,2,3\}$.
    Then a permutation $\pi$ on $I$ and $k \in\{0,1\}$ exists,
    such that
    \begin{eqnarray}\label{eq:lemma.2.1}
      x_{\pi(1),2-k} \cdot x_{\pi(3),1+k} &\clle&   x_{\pi(1),1+k} \cdot
      x_{\pi(3),2-k}
      \\\nonumber
      x_{\pi(2),1+k} \cdot x_{\pi(3),2-k} &\clle&   x_{\pi(2),2-k} \cdot
      x_{\pi(3),1+k}
    .\end{eqnarray}
  \end{lemma}
  \proof{
    Without loss of generality we may assume that
    $$
    x_{12} x_{31} \clle x_{11} x_{32}
    .$$
    If $x_{21} x_{32} \clle x_{22} x_{31}$ choose $\pi=c(1,2,3)$ and $k=0$.
    Otherwise we distinguish two cases:
    (i) if $x_{11} y_{22} \clle x_{12} x_{21}$ then choose $\pi=c(2, 3,
    1)$ and $k=0$ as $
    x_{12} x_{31} \clle x_{11} x_{32}
    $ holds true;
    (ii)  $x_{22} x_{31} \clle x_{21} x_{32} $ and
    $x_{12} x_{21} \clle x_{11} y_{22} $ hold; choose $\pi=(3, 1, 2)$
    and $k=1$ here.
  }

\begin{theorem}\label{thm:2}
  Let  $S$ be a semimodule  over a semifield $\GF$.
  If $V \subset S$ is a subsemimodule  with rank
  equal to $0$ or $1$, then the dimension of $V$ exists.
  If $\GF$ is additionally strictly preordered and  $d \le2$,
  then any subsemimodule  of $\GF^d$ has rank at most $d$.
\end{theorem}

  \proof{
    For $V=\{0\}$ the assertions are
    obvious.  Let $\rk(V)=1$ and let $x \in V$ be a basis vector.
    Assume $v, w\in S$  are two distinct elements of another basis of
    $V$.
    Then $\alpha ,\beta \in \GF\setminus\{0\}$ exist such that $v= \alpha  x$ and $w=\beta x$.
    Hence, $\beta  v = \alpha  w$, a contraction to the independence of $v$ and $w$.

    Let $\rk V =2$ and assume $x=(x_1,x_2),y=(y_1,y_2),z=(z_1,z_2)$ are
    elements of a basis of $V$.
    Lemma \ref{lemma:2}  yields that, without loss of generality,
    \begin{eqnarray}\label{eq:lemma2.appl1}
      x_1 z_2 &=& x_2 z_1 \plus \alpha_{xz} x_1 z_2\\\label{eq:lemma2.appl2}
      y_2 z_1 &=& y_1 z_2 \plus \alpha_{zy} y_2 z_1
    \end{eqnarray}
    for some $a_{xz},a_{zy}\in \GF \setminus\{0\}$.
    Indeed, the value $0$ is excluded due to the independence
    assumption.
    We distinguish three cases: (i) one of the components of $x$, $y$, or $z$ equals~$0$; (ii)
    none of the components equals $0$ and a permutation of $x$, $y$ and $z$ and the
    indices of the component exists such that the conditions
    \eqref{eq:lemma2.appl1}
    and \eqref{eq:lemma2.appl2} hold, but not
    \begin{eqnarray}\label{eq:cond0}
      a_{zy}x_2z_1 \plus\alpha_{xz}x_1 z_2 = 0
    ; \end{eqnarray}
    (iii)
    none of the components equals $0$ and conditions \eqref{eq:lemma2.appl1}--\eqref{eq:cond0}
    holds for some permutation.
    \begin{itemize}
      \item
        Without loss of generality let us assume that $x_1=0$. The assumption
        that $x_2y_1z_1=0$ directly contradicts the assumption of
        semi-linear independence. So we may assume $x_2y_1z_1\not =0$.
        Furthermore we may assume that $z_1 y_2 \clle y_1 z_2$. Then $\alpha  \in
        \GF$ exists such that  $z_1 y_2  \plus  \alpha  y_1 z_2 =  y_1 z_2$.
        Then we have
        \begin{eqnarray*}
          \alpha  y_1 z_2 \left(
          \begin{array}{c}
            0 \\ x_2
          \end{array}
          \right) \plus x_2 z_1  \left(
          \begin{array}{c}
            y_1 \\ y_2
          \end{array}
          \right)
          &=&
          x_2 y_1
          \left(
          \begin{array}{c}
            z_1 \\ z_2
          \end{array}
          \right)
        .\end{eqnarray*}
        This is a contradiction to the assumed independence.

      \item
        Let $\alpha =\alpha_{zy}y_2 z_1$ and $\beta =\alpha_{xz}x_1 z_2$.
        If \eqref{eq:cond0} does not hold then $\alpha  x_1 x_2 \plus  \beta  x_1
        y_2\not=0$ and  we get a contradiction to the independence of $x,y,z$ by
        \begin{eqnarray*}
          (\alpha   x_2 z_1 \plus   \alpha \beta ) \left(
          \begin{array}{c}
            x_1 \\ x_2
          \end{array}
          \right) \plus  \beta  x_1 z_2  \left(
          \begin{array}{c}
            y_1 \\ y_2
          \end{array}
          \right)
          &=&
          (\alpha  x_1 x_2 \plus  \beta  x_1 y_2)
          \left(
          \begin{array}{c}
            z_1 \\ z_2
          \end{array}
          \right).
        \end{eqnarray*}

      \item
        %
        Equality \eqref{eq:cond0} implies
        \begin{eqnarray*}
          \alpha_{zy} z_1\left(
          \begin{array}{c}
            x_1 \\ x_2
          \end{array}
          \right)
          \plus
          \alpha_{xz} x_1\left(
          \begin{array}{c}
            z_1 \\ z_2
          \end{array}
          \right)
          & =&
          x_1 z_1\left(
          \begin{array}{c}
            \alpha_{zy} \plus \alpha_{xz} \\ 0
          \end{array}
          \right)
          \\
          \alpha_{xz} z_2\left(
          \begin{array}{c}
            x_1 \\ x_2
          \end{array}
          \right)
          \plus
          \alpha_{zy} x_2\left(
          \begin{array}{c}
            z_1 \\ z_2
          \end{array}
          \right)
          &  =&
          x_2 z_2\left(
          \begin{array}{c}
            0\\
            \alpha_{zy} \plus \alpha_{xz}
          \end{array}
          \right)
        \end{eqnarray*}
        The assumption $\alpha _{zy} \plus \alpha_{xz}=0$ contradicts the assumption that
        $x$ and $z$ are independent. Hence, $x_1x_2 z_1 z_2 (\alpha _{zy} \plus
        \alpha_{xz}) y\not=0$ and can be represented by a $\plus$-linear
        combination of $x$ and $z$.
      \end{itemize}
      \endproof
    }

Even if $\GF$ is strictly preordered,
a $d$-dimensional subsemimodule  $V\subset \GF^d$ can be a genuine subset.

As usual, $\delta^i$ denotes the $i$-fold product of $\delta\in \GF$.
\begin{theorem} \label{thm:3}
  \felix{\marginpar{Martin!\\\"uberpr\"ufen!}}
  Let $S = \GF^3$ be a semimodule  over a semifield $\GF$
  and
  $n\in \NN\cup \{\infty\}$.
  Assume that $\delta \in \GF$ exists such
  that
  the following condition holds true:
  \begin{eqnarray}
    \label{eq:def.cond.1}
    \gamma \delta ^i &\cg& \gamma  \plus \beta \quad \text{ for all }1 \le i < n + 1,  \gamma ,\beta \in
    \GF, \gamma  \not = 0,
    \beta \not\cge \gamma
  \end{eqnarray}
  Then a  subsemimodule  $V \subset S$ exists with $\rk V \ge n$.
\end{theorem}

Roughly speaking, condition \eqref{eq:def.cond.1} requires the
existence of a number $\delta \cg
1 \plus1 $. This is true for the maximum semifield
$([0,\infty), \vee, \cdot)$, where \eqref{eq:def.cond.1} holds for
all $n\in\NN$.
\felix{Proposition \ref{xxxx} shows that the rank of $V \subset \GF^3$ can
even be uncountably
infinite.}
If $\GF$ is a proper semifield, then condition
\eqref{eq:def.cond.1} is void.

\proof{
  We may assume that $n$ is finite and define $x_i = (1, \delta ^i, \delta ^{2i})$ for $1\le i \le n$.
  In order to show that
  $$
  \dotsum_{i\not = j}\alpha_i x_i \not= \gamma  x_j
  $$
  for all $1\le j \le n$, $\alpha _i \in \GF$ and $\gamma  \in \GF\setminus\{0 \}$,
  it suffices to show  that for all $1\le j \le n$ and $\gamma \in \GF\setminus\{0\}$ the two equalities
  \begin{eqnarray}\label{eq:dim.ann.1}
    \dotsum_{i \not = j}\alpha_i & =& \gamma
    \\\label{eq:dim.ann.2}
    \dotsum_{i \not = j}\alpha_i\delta ^i  &=& \gamma  \delta ^j
  \end{eqnarray}
  imply
  \begin{eqnarray}                                   \label{eq:def.contra}
    \dotsum_{i > j}\alpha_i \delta ^{2i} &\not \cle& \gamma  \delta ^{2j}
                                                     .
  \end{eqnarray}
  Note that \eqref{eq:def.cond.1} implies
  \begin{eqnarray}\label{eq:def.rule.1}
    \delta ^j \cg \delta ^i \text{ for  }0\le i\le j < n+1
    .    \end{eqnarray}
  Definition \ref{def:inferior} immediately yields
  \begin{eqnarray}\label{eq:def.rule.2}
    \alpha  \cle \beta  &\Rightarrow& a \gamma  \cle \beta \gamma  \text{ for all } \gamma \in \GF
    \\ \label{eq:def.rule.3}
    \alpha  \cle \beta , \gamma  \cle d &\Rightarrow &a \plus \gamma  \cle \beta  \plus d \text{ for all }
                                                       \gamma \in
                                                       \GF
  \end{eqnarray}
  Now assume that $\dotsum_{i > j}\alpha_i \delta ^{2i} \cle \gamma  \delta ^{2j}$.
  Then equality \eqref{eq:dim.ann.2} and calculation rules
  \eqref{eq:def.rule.1}--\eqref{eq:def.rule.3} yield
  \begin{eqnarray*}
    \gamma  \delta ^{2 j + 1}
    & = &  \delta ^{j + 1}\left(\dotsum_{i > j}\alpha_i \delta ^i \plus \dotsum_{i < j}\alpha_i \delta ^i\right)
    \\& \cle &  \dotsum_{i > j}\alpha_i \delta ^{2i} \plus   \dotsum_{i < j}\alpha_i  \delta ^{i+ j +1}
               \cle \gamma  \delta ^{2j} \plus  \delta ^{2j} \dotsum_{i < j}\alpha_i
               \cle \left(\gamma \plus \gamma \right) \delta ^{2j}
  \end{eqnarray*}
  In total, this chain of inequalities contradicts condition
  \eqref{eq:def.cond.1}, so that  $\dotsum_{i > j}\alpha_i \delta ^{2i} \not\cle
  \gamma  \delta ^{2j}$ holds true.

}

\subsection{Rank of  a matrix}\label{sec:matrixrang}
  Let $H:\GF^d\rightarrow \GF^d$ be a $\plus$-linear map.
  From the point of view of a PCA, we expect that
   $H$ does not enter into a linearly inferable ``$p$-dimensional'' $\PCA_p$, if
  $\im H \not \subset \Span\{b_1,\ldots, b_p\}$ for all vectors
  $b_1,\ldots, b_p \in \GF^d$.
  This basic requirement is ensured by the definition of the set $B$ in
  Definition \ref{def:PCA} of a PCA.

  The unique rank of a matrix in standard linear algebra
  splits into different notions 
  already in the tropical algebra \citep{akian2006max,maclagansturmfels}.
   We call a matrix $M \in\GF^{k\times d}$ tropical, if $\GF =
  ((0,\infty),\vee,\cdot)$.
\cite{akian2006max} and  \cite{guterman2016rank} 
  give overviews over various
  rank definition for a tropical matrix, most of them can be
  immediately generalized to arbitrary semirings $\GF $.
  For instance, the row rank and the column rank can be defined using
  any of the definitions of independence. A stronger form of Definition
  \ref{def:indep} leads to ranks in the so-called Gondran–Minoux sense.
  The so-called tropical rank is frequently used. Its definition is more
  complex, but the tropical rank can be interpreted as the dimension
  of the tropical linear span \citep{guterman2016rank}.
  The Kapranov rank has a convenient interpretation of the least
  dimension of a tropical linear space that includes the rows
  (columns) of $M$.
  The  Barvinok rank, or factor rank, is the smallest $r$, such that
  $$
  M = H_1 H_2^\top, \qquad H_1 \in R^{k \times r}, H_2 \in R^{r \times d}
  .$$
  Various relations between the different notions of a rank exist.
For example, the tropical rank is less than or equal to the
Kapranov rank,
which is itself less than or equal to the  Barvinok rank.
  However, they do not bound each other, in general  \citep{kimroush06}.

  The determination of the rank of a matrix is typically an NP hard
  problem, see \cite{guterman2016rank}. In some cases, it is even
  undecidable \citep{kimroush06}.
  The problem to determine whether the rank is less than or equal to a
  given number, is, in general, polynomial, i.e., much simpler.
  The construction in Remark
  \ref{rem:barvinok} even guarantees
  immediately that the 
  Barvinok rank is at most $p$.

    Note that our
  approach of a PCA  is different from the so-called tropical PCA, which is a
  purely geometrical approach leading to a different optimization problem
  \citep{page2020tropical}. 

\end{appendix}

  \bibliographystyle{plainnat} 


\end{document}